# Sliding Window Regression based Short-Term Load Forecasting of a Multi-Area Power System

Irfan Ahmad Khan
*Marine Engineering Tech. Department*
*Texas A&M University at Galveston*
Galveston TX USA
irfankhan@tamu.edu

Adnan Akbar
*Department of Informatics*
*Kings College London*
London UK
adnan.akbar@kcl.ac.uk

Yinliang Xu
*Environment Science and New Energy Tech.*
*Tsinghua-Berkeley Shenzen Institute*
*Shenzen P.R. China*
xu.yinliang@sz.tsinghua.edu.cn

***Abstract*—Short term load forecasting has an important medium for a reliable, economical and efficient operation of power system. Most of the existing forecasting approaches utilize fixed statistical models with large historical data for training the models. However, due to recent integration of massive distributed generation, nature of load demand has become dynamic. Thus because of dynamic nature of the power load demand, performance of these models may deteriorate over time. To accommodate the dynamic nature of the load demands, we propose sliding window regression based dynamic model to predict the load demands of multi-area power system. Proposed algorithm is tested on five zones of New York ISO. Results from our proposed algorithm are compared with four existing techniques to validate the performance superiority of the proposed algorithm.**

## I. Introduction

Load forecasting has become an important factor for a reliable and economical operation of power systems. Depending on the time horizon, load forecasting can usually be classified into short-term, midterm-term load forecast (MTLF), and long-term. Ranging from an hour to a week, short-term load forecasting (STLF), is essential for many functions such as unit commitment, economic dispatch [1], energy transfer scheduling and real-time operation and control [2,3]. Covering from a few weeks to several years, mid and long-term load forecasting is used for maintenance scheduling, adequacy assessment, purchasing fuel, scheduling of fuel supplies and limited energy resources, etc. [4].

Accurately estimated forecasts are essential part of the electricity utility's operation and production costs. Overestimation of electricity load demand will lead to the excessive energy purchase or start-up of too many units, thereby supplying an unnecessary level of reserve. Underestimation, on the other hand, may result in a risky operation, with insufficient level of spinning reserve, causing the system to operate in a vulnerable state to the disturbance [5, 6]. Therefore, a wide variety of forecasting models have been proposed, most of which can be generally classified into two broad categories: statistical methods and artificial intelligence (AI)-based methods. Most statistical models based on linear analysis have deficiencies in solving the load forecasting problem, because the load series are usually nonlinear. In recent years, AI-based techniques such as neural networks have been very popular in finding promising results.

## II. Background

Usually, modeling of a regression problem is performed by three ways. The traditional approach to model regression problem is by using statistical methods like autoregressive integrated moving average (ARIMA) model which breaks the time series into different components e.g. trend components and seasonality components and estimates a model for each component. However, it requires an expert in statistics to calibrate its model parameters [7]. Another approach to model the historical data is to devise a list of temporal features so that the auto correlation information is not lost. Some of the most commonly used temporal features are the time since certain event, time between two events, and entropy measurements etc. Afterwards, techniques like Random Forest and Gradient Boosting are applied on these features to observe relative feature importance [8]. By doing so, we can keep healthy features and drop the ineffective features. The third approach is to use the Sliding Windows Regression that is an interesting prediction technique and can provides impressive results without much prior experience. While predicting next value $x$(t+1), the idea is to feed not only $x$(t), but also $x$(t-1), $x$(t-2) etc. to the model. In this way, it incorporates auto correlation information into the model [9].

Load forecasting is usually made by constructing models based on the historical load demand data and climate change etc. [10]. Support Vector Machine (SVM) was first time applied to load forecasting in [10] where it was observed that Support Vector Regression (SVR) performs well for time series analysis. SVR is an extension of SVM which is used for regression analysis. The main idea of SVR is same as SVM that maps the training data into higher feature space using kernel functions and finds the optimum function that best fits the training data. Furthermore, a modified version of the SVR was proposed to solve the load forecasting problem in [11] where the risk function of the SVR algorithm is modified with the use of locally weighted regression (LWR) while keeping the regularization term in its original form. In addition to it, two improvements to the SVR based STLF method: procedure for generation of model inputs and subsequent model input selection were introduced in [12] using feature selection algorithms. Similarly, SVM was used in [13] where a hybrid model was proposed to forecast the responses of the controlled thermal loads and forecasting the residual. In a power system covering a large geographical area, a single model for load forecasting of the entire area sometimes may not guarantee satisfactory forecasting accuracy. One of the major reasons is because of the load diversity, usually caused by weather diversity, throughout the area. Multi-region load forecasting will be an effective solution to generate more accurate forecasting results, as well as provide regional forecasts for the utilities. A SVR based multi-area load forecasting system for day-ahead operation and market is proposed is [14]. However, it was not considered for shorter interval of time which is more important especially in a system with huge amount of stochastic distributed generation.

Most of the existing methods use fixed models with large amounts of historical data for training. However, performance of these models may deteriorate over time as the statistical properties of the underlying data may change with time due to concept drift, especially for the case of power system load data which is dynamic. But, model remains unchanged due to large amount of historical data to train the model. [15] concludes that most of existing approaches for STLF are not applicable on local load forecasting due to long training time. More recent time series prediction methods address these issues [16] where different variants of moving window are suggested. However, most of these methods depend upon individual application and lack a generic solution for applying to different domains.

We propose an adaptive prediction algorithm called Sliding Window Regression (SWR) for dynamic load data prediction. Generally, prediction models are trained with large historical data and once the model is trained it may not be updated due to limitations posed by large training time and thus, such models may not be optimized to perform under concept drift [17]. The context of the application or the real data may change resulting in the degraded performance by the prediction model. For such scenarios like real time load data prediction, we develop a prediction model which utilizes sliding window of data for training the model; and once new data arrives, it calculates an error and incorporates it in the model accordingly.

As power system load data is dynamic; thus, we forecast the short-term load using SVR and training the model using sliding window. We called it Sliding Window Regression. For each time interval, we automatically find the optimal window size for the training data using the Lomb Scargle method to find the optimum size of training window [18]. Our proposed approach is adaptive in nature as it tracks down errors and prevents it from propagating by retraining the model periodically. The size of the prediction window or forecast horizon is also adaptive and is derived by the performance of the model in order to ensure a certain reliability in the prediction.

There are several loss functions which describe the performance of the prediction model e.g. Mean absolute error (MAE), Root Mean Square Error (RMSE), Mean Absolute Percentage Error (MAPE), Root Mean Square Percentage Error (RMSEP) and Almost correct Predictions Error rate (ACPER). In this paper, we have used MAPE which shows the relative accuracy of the regression problem as follows:

$$\text{MAPE} = 100\frac{\sum_{i=1}^{n}\left|\frac{L_i - \hat{L}_i}{L_i}\right|}{n} \quad (1)$$

where $L_i$ is the predicted value and $\hat{L}_i$ is the actual load value and n is the size of training window.

## III. Proposed Algorithm

Sliding Window Regression based proposed algorithm consists of three main steps: selection of regression algorithm, finding optimal training window size and finding the size of prediction window/horizon as shown in Fig.1. There are several algorithms available in literature for time series prediction ranging from statistical to pure ML domain algorithms. We have adopted SVR due to its ability to model no-linear data using kernel functions.

SVR algorithms provide more accurate models as their counterparts at the expense of additional complexity. However, as in our algorithm we propose to use a small training window, the added complexity is almost negligible for such small datasets. Choice of optimal training window size is an open research area for Machine Learning models. Generally, accuracy of prediction model improves as the size of the training window increases i.e. it is better to have large historical data for training the prediction model so that it covers all possible patterns spanning time series. However, there is a major drawback of having large training data: if the behavior of the underlying model changes, trained model may not track the changes and result into erroneous readings.

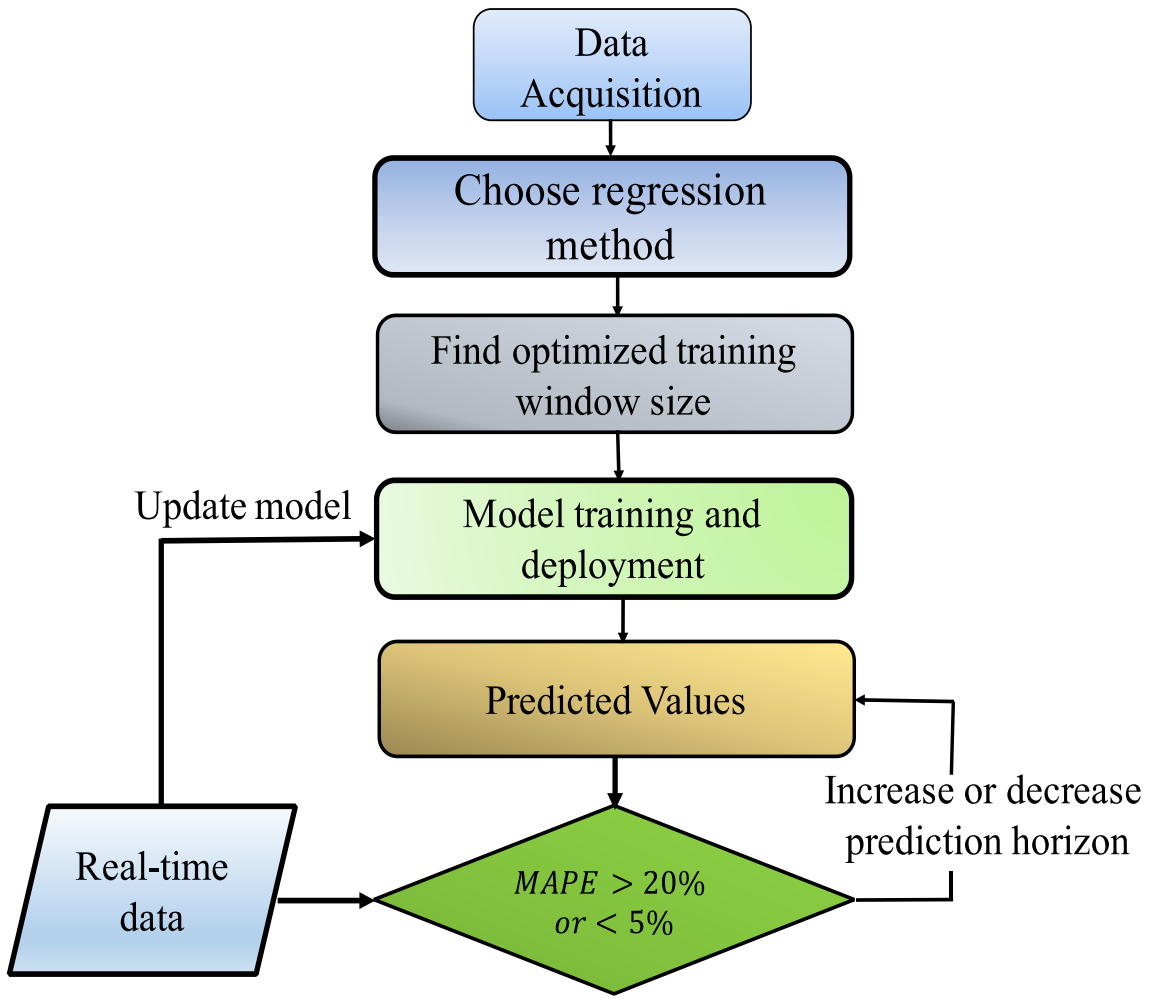

Fig. 1 Flow chart for proposed SWR algorithm

In contrast to it, we have proposed to use the sliding window for training the model in which the most recent data will be fed into the model. We have proposed a novel method based on Lomb Scargle method to find the optimum window size by exploiting the inherent periodic nature of the power load demand time series data and validated our results by comparing it with other existing prediction techniques.

Finding an optimum size of prediction window or more commonly known as prediction horizon is very important to ensure a certain level of accuracy. The idea behind finding an optimum prediction horizon is to increase the size if the accuracy of the model is high and decrease it if accuracy is low than a certain level as shown in Algorithm I.

**Algorithm I** Adaptive Prediction Window Size

```
function PREDICTIONWINDOW (y_act, y_pred)
… MAPE=mean(abs((y_act - y_pred)/y_act) * 100)
..  if    MAPE>20%        then
..  PredictionWindow = PredictionWindow -1
..  else if MAPE<5%       then
..  PredictionWindow=PredictionWindow +1
..  else
..  PredictionWindow = PredictionWindow
..  end if
..  return PredictionWindow
end function
```

## IV. RESULTS

Proposed rolling window regression algorithm is simulated for five zones of New York ISO: 1) Capital Zone C, 2) Central Zone C, 3) Dunwodie Zone I, 4) Genesee Zone B and 5) Valley Zone G. Load data of these 5 zones is collected for two weeks starting from October 16, 2017 at midnight to October 29, 2017 at 13:30 pm with each reading after 5 minutes that makes total number of readings equal to 3906 [19].

It is obvious from Fig. (2-6) that the predicted value follows the actual value with small error. Error between the predicted value and actual value has been calculated in terms of MAPE and shown in Fig. 7. Our proposed algorithm is compared with four existing techniques of Linear Regression, SVM with RBF kernel, Decision tree regression, and random forest regression as shown in Fig. 7.

We have compared the performance of several variants of SVR and finally SVR with adaptive sliding window for training dataset and adaptive prediction horizon was chosen as underlying regression algorithm. It is clear that our proposed algorithm shows 2% of MAPE which is less than that of other existing four regression techniques. It validates the effectiveness of the proposed algorithm.

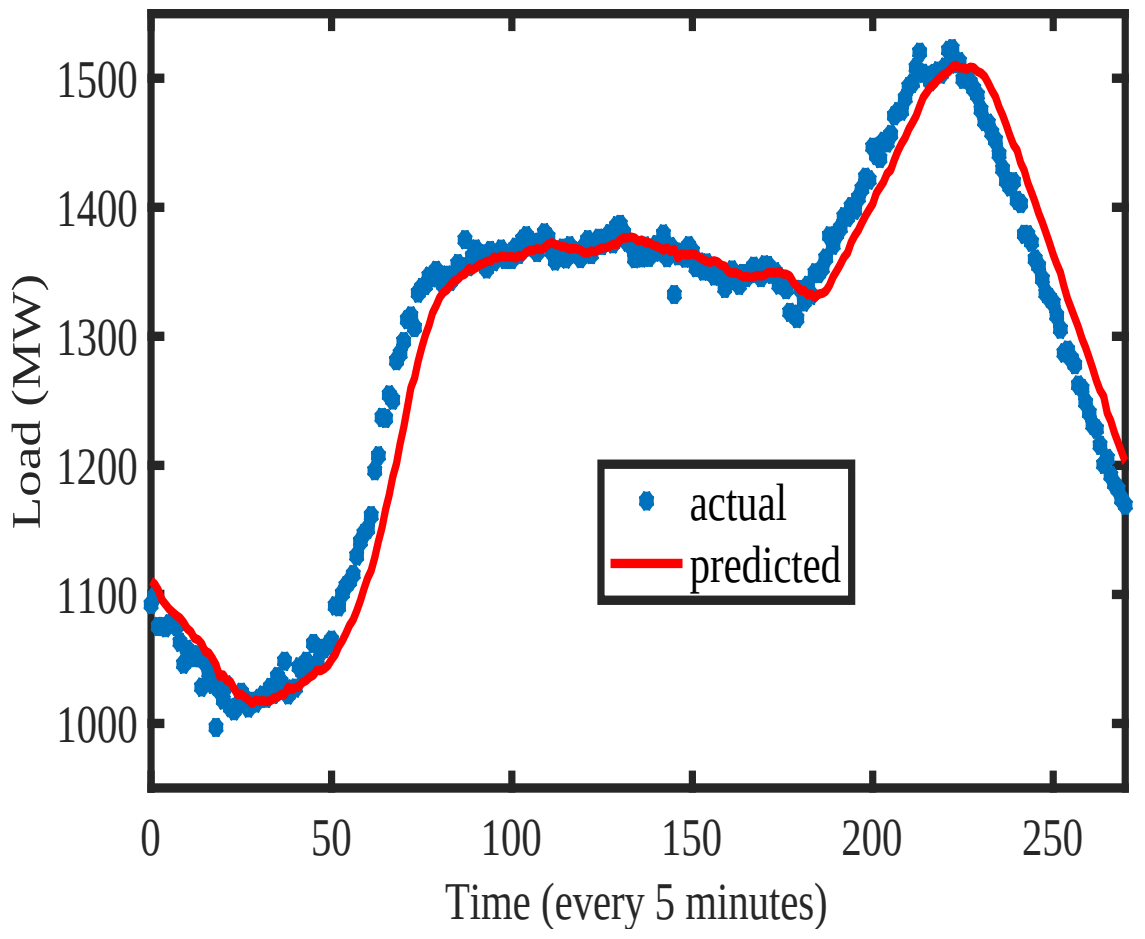

Fig. 2 Actual and predicted load forecast of Capital Zone

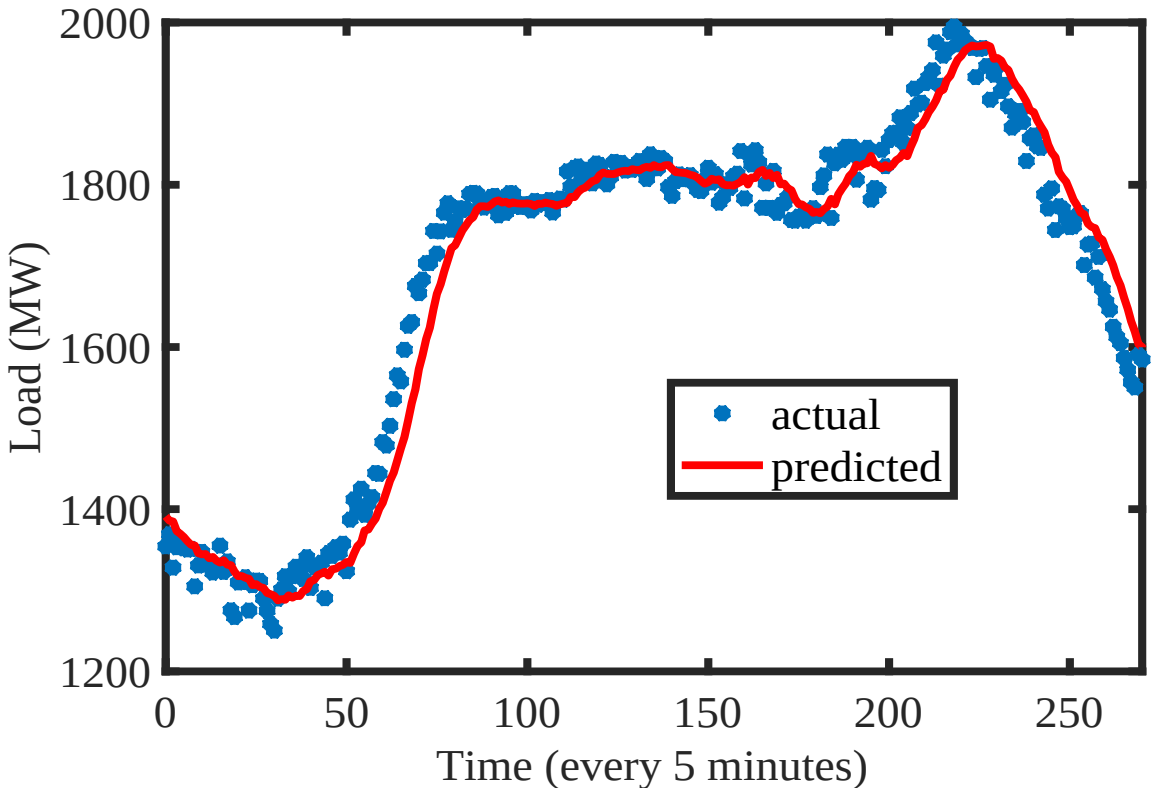

Fig. 3 Actual and predicted load forecast of Central Zone

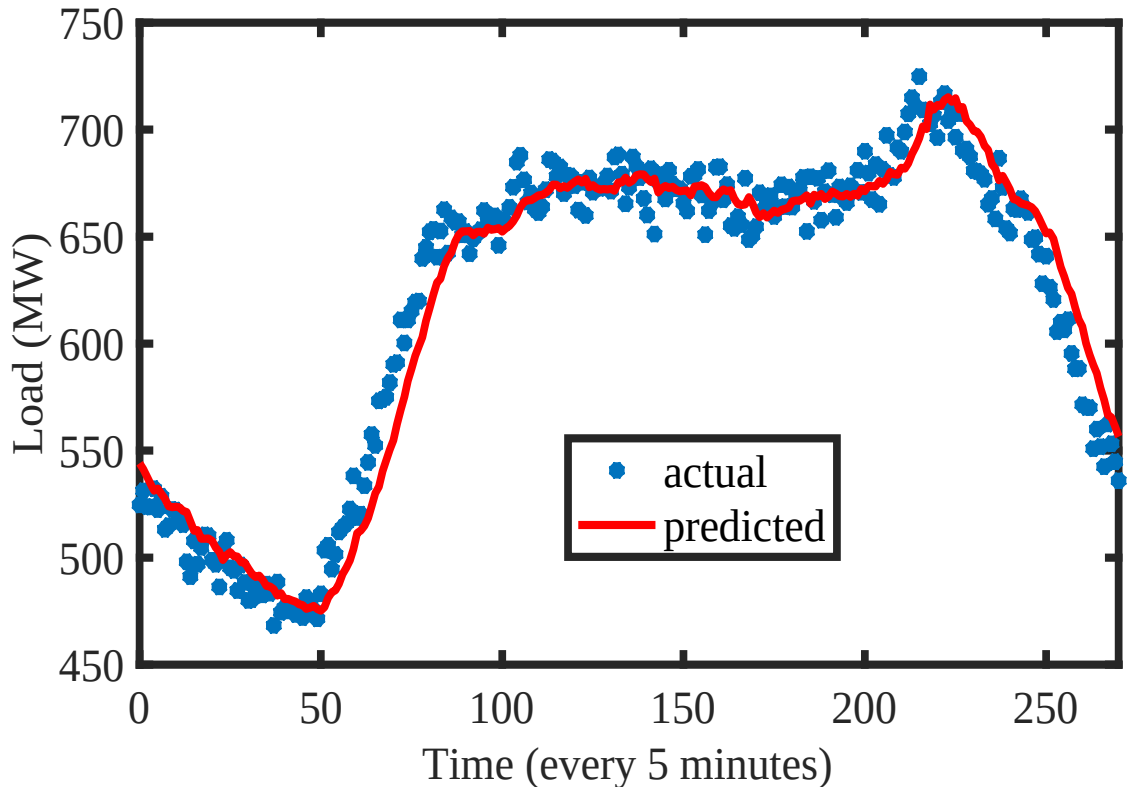

Fig. 4 Actual and predicted load forecast of Dunwodie Zone

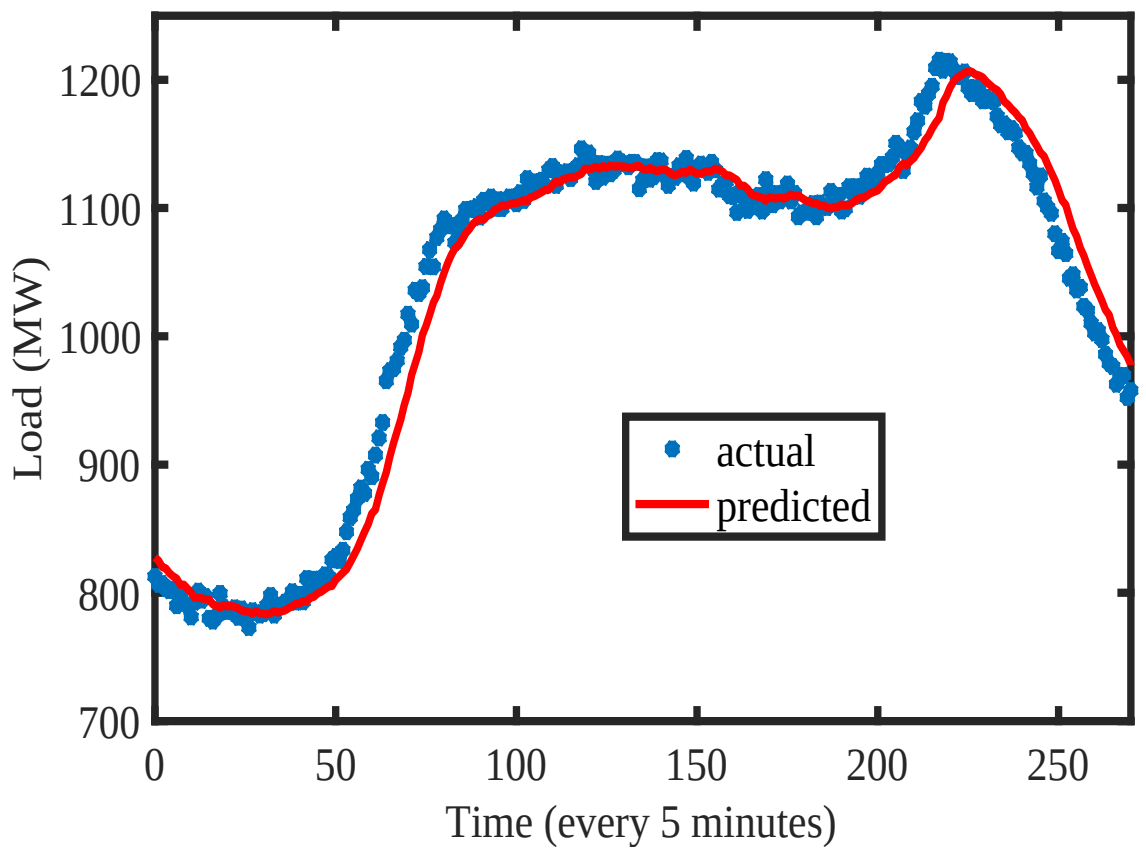

Fig. 5 Actual and predicted load forecast of Genesee Zone

## V. CONCLUSION

In this paper, we solved the problem of power load forecasting of multi area power system. To incorporate the dynamic nature of power load demand, we have proposed a sliding window to train the regression model. We have also proposed to find an optimum size of prediction horizon to improve the accuracy of regression model. The proposed algorithm has been tested on multi-area power system of New York ISO where we have selected 5 zones to test of algorithm. Simulation results are then compared with various regression techniques of Linear Regression, SVM with RBF kernel, Decision tree regression, and random forest regression. Simulation results show that our proposed algorithm forecast the load demand data with less MAPE error than the existing algorithms in terms of percentage error of MAPE.

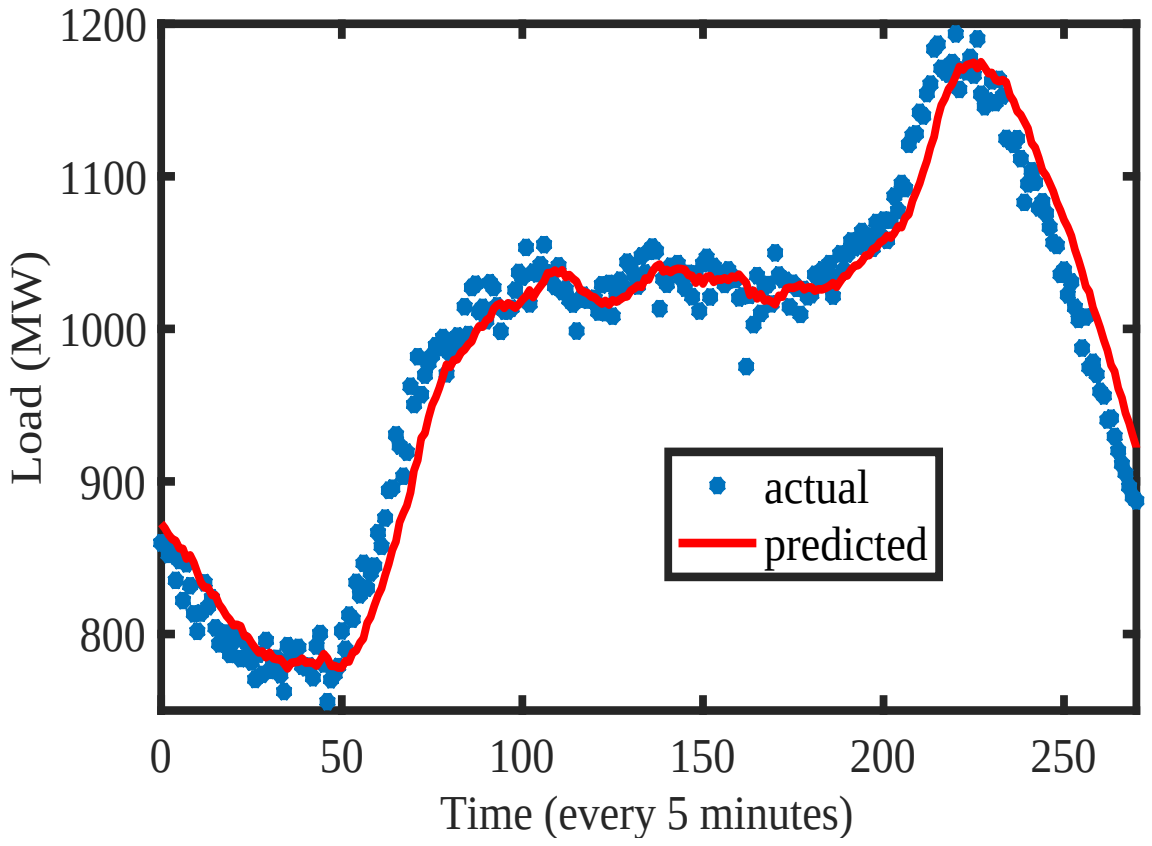

Fig. 6 Actual and predicted load forecast of Valley Zone

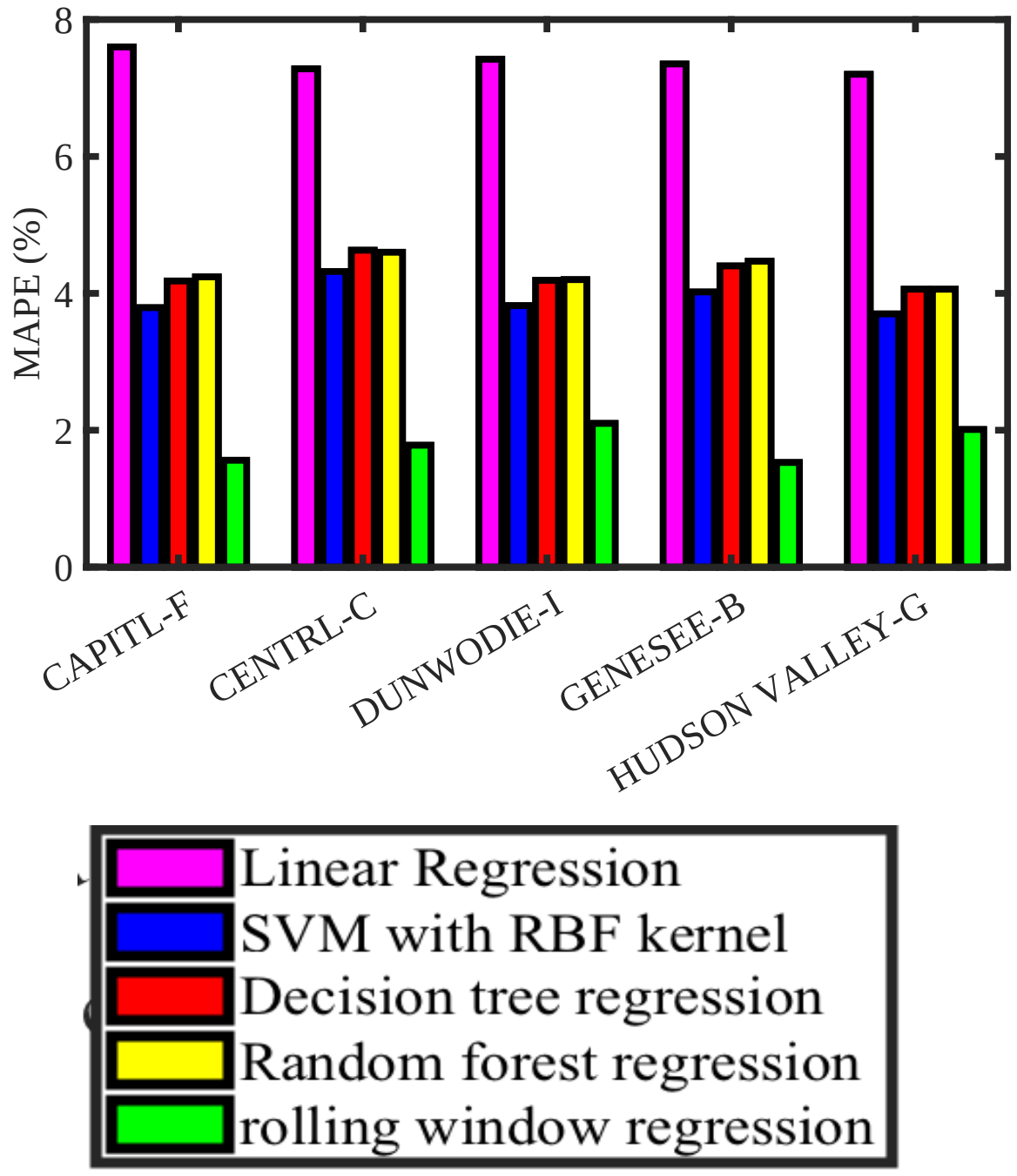

Fig. 7 Comparison of proposed algorithm with existing forecasting schemes in terms of MAPE.